\documentclass[twocolumn,showpacs,superscriptaddress,preprintnumbers,amsmath,
amssymb,prb]{revtex4}

\usepackage{graphicx}
\usepackage{dcolumn}
\usepackage{bm}
\usepackage{hyperref}

\begin{document} 

\title{Effects of Sr doping on the magnetic properties of Sr$_x$Ba$_{1-x}$CoO$_{3}$
}

\author{V. Pardo}
 \email{vpardo@usc.es}
\affiliation{
Departamento de F\'{\i}sica Aplicada, Facultad de F\'{\i}sica, Universidad
de Santiago de Compostela, E-15782 Campus Sur s/n, Santiago de Compostela,
Spain
}
\affiliation{
Instituto de Investigaciones Tecnol\'ogicas, Universidad de Santiago de
Compostela, E-15782, Santiago de Compostela, Spain
}

\author{J. Rivas}
\affiliation{
Departamento de F\'{\i}sica Aplicada, Facultad de F\'{\i}sica, Universidad
de Santiago de Compostela, E-15782 Campus Sur s/n, Santiago de Compostela,
Spain
}

\author{D. Baldomir}
\affiliation{
Departamento de F\'{\i}sica Aplicada, Facultad de F\'{\i}sica, Universidad
de Santiago de Compostela, E-15782 Campus Sur s/n, Santiago de Compostela,
Spain
}
\affiliation{
Instituto de Investigaciones Tecnol\'ogicas, Universidad de Santiago de
Compostela, E-15782, Santiago de Compostela, Spain
}

\begin{abstract}

Magnetic properties of the Sr-doped BaCoO$_3$ are explained on the basis of
\textsl{ab initio} calculations and the analysis of experimental
literature. Formation of magnetic clusters bigger than in the parent
compound and increase in the blocking temperature are observed.
Superparamagnetism remains at room temperature. The possibility of
tuning the size and properties of the magnetic clusters with doping is explored.

\end{abstract}

\pacs{71.15.Ap, 71.15.Mb, 71.20.Be, 75.20.-g}

\maketitle

\textsl{Ab initio} electronic structure calculations are becoming
increasingly important in materials science. The improvement of the codes
and the ever increasing capabilities of modern supercomputers has produced
a comparable improvement in the precision of the properties of solids
which are attainable nowadays by means of, e.g., density functional theory
calculations\cite{dft}. The connection with the real physical properties of the
materials is already possible and the project of designing materials thoretically 
has been under way
for the last few years\cite{apl_design}.

Transition metal oxides have drawn the attention of the scientific
community for the last 50 years
because of their interesting physical properties and multiple
applications\cite{review_tmo}. Co oxides are becoming increasingly important,
mainly because of their thermoelectric\cite{thermo},
magnetoresistive\cite{co_cmr} and superconducting\cite{co_sc} properties.
These materials also present difficulties for the \textsl{ab initio}
calculations because of having strongly correlated electrons; the use of
the LDA+U method\cite{anisimov} or dynamical mean field theory\cite{dmft}
is needed to predict their properties correctly.

Systems formed by magnetic nanostructures have been studied intensively in the
last few years, showing some interesting technological
applications, in fields such as magnetic recording, sensors, MRAM and
magnetoelectronics\cite{review_batlle,applications}.
E.g., Co nanocrystals have also numerous applications due to
their special magnetic properties\cite{apl_coparticles}.

In a recent paper\cite{vpardo_clusters}, the magnetic properties of BaCoO$_3$,
were analyzed considering the formation of nanometric magnetic clusters,
ferromagnetic regions embedded in a non-ferromagnetic matrix, in the
material, starting from experimental data and interpreting them by means
of \textsl{ab initio} calculations.
Glassy behavior has also been recently found in similar
compounds\cite{tona,mira97,spinglass,mira01,jmmm_glass,rivas}. This is usually related to
having a sort of self-generated assembly of magnetic clusters in which
magnetic interactions introduce glassiness among
them\cite{rivadulla_prl,rivas_physb} or a
competition between ferromagnetic and antiferromagnetic couplings which
may lead to frustration of any long range magnetic order, as it is the case in
BaCoO$_3$\cite{vpardo_prb}.

In this paper, we will discuss how these magnetic clusters change with
Sr-doping, focusing in the half-doped compound Sr$_{0.5}$Ba$_{0.5}$CoO$_3$.
The description of its magnetic properties will be made using the
experimental data available, interpreted via \textsl{ab initio}
calculations. 


The temperature and applied field dependence of the magnetization
measured after cooling in zero field (ZFC) and after cooling in a field
(FC) in Ref. \onlinecite{yamaura} (Fig. 12) suggest us that the Sr-half-doped
compound behaves as an assembly of non-strongly interacting particles,
which we proceed to analyze. 
Two cusps appear in the ZFC magnetization curves, one at 50 K and another one at about 140
K, being the cusp at about 50 K a remainder of the parent compound which
comes out in the doped case because of the fabrication procedure.
The physics which is genuine of the doped compound is shown by the
higher temperature one. 
This so-called bimodality, i.e., the appearance of two types of magnetic
particles of different sizes in one sample, is usually found in granular systems 
\cite{bimodality,apl_coparticles}. Similar curves could also be understood
taking into account the 
effects of the surface of the clusters\cite{prb_surface,fiorani_surf}
but we have ruled this possibility out because
of the technique used to produce the sample, which would tend to conserve
a matrix with the properties of the parent compound.

For describing properly what the magnetic morphology of the system is we
need to characterize the size of these magnetic clusters and the
interparticle distance. For doing so, we
will make use of \textsl{ab initio} calculations performed with the WIEN2k
software\cite{wien}, which uses an APW+lo full-potential method\cite{sjo}
and the LDA+U approach\cite{sic} in the ``fully-localized
limit"\cite{mazin}, to deal with the strong correlations typically present
in transition metal oxides. 

\begin{figure}
\includegraphics{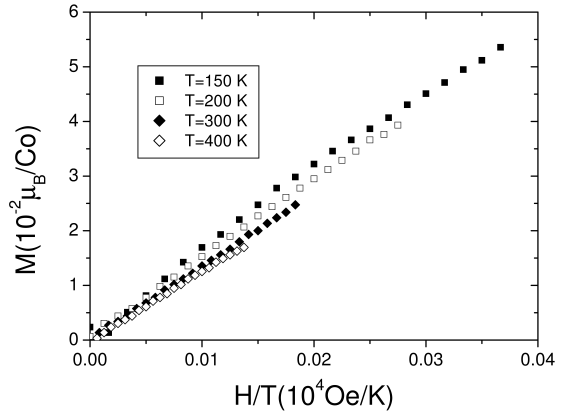}
\caption{Total magnetization vs. H/T of Sr$_{0.5}$Ba$_{0.5}$CoO$_3$ for
several temperatures (experimental data taken
from Yamaura et al.\cite{yamaura}). The curves superpose as expected for a 
system of superparamagnetic particles above the blocking
temperature. Even at room temperature, superparamagnetism is present.}\label{curve}
\end{figure}

The blocking temperature of Sr$_{0.5}$Ba$_{0.5}$CoO$_3$ is, as can be seen in
Ref. \onlinecite{yamaura}, Fig. 12, at about 140 K. With this value, the size
of the magnetic clusters which become blocked below that temperature, but
act as superparamagnetic particles above it (see Fig. \ref{curve}), can be
estimated with the knowledge of the anisotropy constant of the material.
This quantity can be calculated \textsl{ab initio} just by computing the
total energy of the material, including spin-orbit effects in a second
variational manner\cite{singh}, assuming the
magnetic moments lie in the different crystallographic directions. We
calculated the anisotropy constant as $K=\frac{E_B}{4}$ (we assume cubic
anisotropy due to the nearly octahedral environment where the Co ions are), 
where the magnetocrystalline energy $E_B$ is defined as the work required
to make the magnetization lie along the hard ($a$) axis compared to an
easy (c) direction.
We must note here that the introduction of Sr produces severe changes in the
crystalline anisotropy energy of the material. It changes the easy axis
from the hexagonal plane where it lies in the parent compound to the
Co-chains axis (c). Also, the magnitude of the anisotropy constant has in
the doped compound a value of $1.5\times10^7$erg/cm$^3$, which is one
order of magnitude smaller than in the parent compound, but still larger 
than the typical values for
systems of Co magnetic particles, hence, still one can assume that the
formation of the clusters will have a much smaller contribution to the
magnetic anistropy of the system and consider that the magnetocrystalline
anisotropy is the dominant effect. 

The size of the clusters can then be obtained using the well-known
expression for systems of superparamgnetic particles\cite{morrish}: $KV=
25k_BT_B$,
where $K$ is the anisotropy constant, $V$ is the typical volume of the
particles, $T_B$ is the blocking temperature (which is about 140 K in this 
case), $k_B$ is the Boltzmann's constant and the factor 25 appears due to the experimental technique. The
typical volume of the clusters which comes out of our calculations is
about 4 nm in diameter, comprising some 500 Co atoms. The size of
these clusters is much bigger than those forming the parent
compound, which are about 1.2 nm in diameter\cite{vpardo_clusters}, but
these are still present in the doped sample utilized in Ref.
\onlinecite{yamaura}, accompanied of the bigger ones due to the
half-doped material. These determine the magnetic properties of the
compound above some 100 K and also at room temperature, where
superparamagnetism remains for the case of the Sr-doped compound, whereas
in the parent compound, the breaking of the clusters occurs at about 250 K
and paramagnetism is the phase at room temperature\cite{yamaura}. Below 100 K, the
properties are determined by the remaining of the smaller clusters, which
become blocked below approximately 50 K. 
In the half-doped compound,  magnetic
clusters containing about 500 Co atoms exist at room temperature.
The possibility of controlling the size and
properties of the magnetic particles with doping is clear. Introducing Sr
into the undoped BaCoO$_3$ produces the appearance of bigger clusters
which can be observed already for the compound Sr$_{0.2}$Ba$_{0.8}$CoO$_3$
(see in
Ref. \onlinecite{yamaura}, Fig. 11 how a second cusp starts to develop in
the ZFC curve at
a higher temperature). The control of the size and magnetic properties of
the clusters can also be done by changing the magnetic
field\cite{rivadulla_prl}, but here we propose a different way to do so, i.e., doping the
sample with an isovalent cation. Moreover, here we are dealing with magnetic particles which
spontaneously occur in the system and whose properties can be tuned by
changing the dopant concentration.

The formation of magnetic clusters in the series
Sr$_x$Ba$_{1-x}$CoO$_3$ ($x\leq0.5$) is probably related to the degeneracy
of both ferromagnetic and antiferromagnetic
configurations\cite{review_spinglass}. Our calculations on the half-doped
compound yield a similar behavior to that found in the parent
compound\cite{vpardo_prb}, ferromagnetism is marginally more stable but it
is orbital ordering the main energetic contribution, so-called alternating
orbital ordering along the Co chains\cite{vpardo_prb} is about 50 meV/Co more stable than the
``ferro-orbital" solution, whereas energy differences
due to magnetic ordering are one order of magnitude smaller. The interplay
between double exchange favoring ferromagnetism and superexchange favoring
antiferromagnetism might porduce frustration in this type of compounds
and a cluster-glass behavior could turn out as a result. However, this
does not seem the case from the susceptibility curves we can see in Ref.
\onlinecite{yamaura}, which do not show strong interactions between the
clusters.

In Fig. \ref{curve}, we plot the data of magnetization vs. H/T which
appears in Ref. \onlinecite{yamaura}, (inset of Fig. 12) as several curves of
magnetization vs. magnetic field for different temperatures. The curves
superpose as expected for a system of superparamagnetic particles above
the blocking temperature. As 
mentioned above, superposition is maintained for the curve at 400 K,
indicating that, at room temperature and above, magnetic clusters still exist and
determine the magnetic properties of the material. At high temperatures,
the slope grows because the clusters get bigger (a lower anisotropy
constant 
when temperature increases implies higher volume of the clusters).

These curves can be fit to a Langevin formula, which approximates
to a Curie law at small values of the ratio $\frac{\mu H}{k_BT}$, where
$\mu$ is the magnetic moment of the cluster, in the following way:
$M \simeq N \frac{\mu^{2}H}{3k_BT}$,
where $N$ is the density of clusters in the material.
From this fitting of the curves in Fig. \ref{curve}, one can estimate the
density of clusters $N$, which determines the way these clusters interact
and also the shape of the FC curve below the blocking
temperature. The result is that there is one cluster formed by about 500
Co atoms in a spherical volume with a diameter of approximately 8 nm. Hence, the average distance between
clusters is about 2 diameters. This means that the clusters do not
strongly interact with each other, as it becomes clear from Ref.
\onlinecite{yamaura}, Fig. 12, where it can be observed that the FC curve
below 140 K (the blocking temperature) rises as expected for
non-strongly-interacting particles\cite{fiorani}, even though this shape might depend on
the cooling speed\cite{cooling_speed}. Cluster-glass behavior is not
observed in the sample and this is confirmed by the calculations showing
that the clusters are not strongly interacting (they are far enough from
each other).

In this paper we have presented an explanation for the magnetic
properties of the series Sr$_x$Ba$_{1-x}$CoO$_3$ characterized by means
of \textsl{ab initio} calculations. The presence of the clusters is an
essential ingredient in the description of the magnetic properties of
the material. When Ba is
substituted by Sr, the magnetic clusters increase in size and
superparamagnetism remains at room temperature for the half-doped compound. The
possibility of tuning the magnetic properties of a system of
superparamagnetic clusters by doping with an isovalent cation is explored. It would be
interesting to analyze what happens to the system when doping it with a
different divalent cation such as Ca. 
The reasoning we follow in this letter might be extended to other similar
compounds showing phase segregation, which are of key importance for
nanoscience and nanotechnology. We state the possibility of studying these
variations in magnetic properties by means of electronic structure
\textsl{ab initio} calculations.

The authors wish to thank the CESGA (Centro de Supercomputaci\'{o}n de
Galicia) for the computing facilities, the Xunta de Galicia for the
finantial support through a grant and the project No. PGIDIT02TMT20601PR
and the Ministerio de Educaci\'on y
Ciencia of Spain by Project No. MAT2004-05130.


\begin{thebibliography}{34}
\expandafter\ifx\csname natexlab\endcsname\relax\def\natexlab#1{#1}\fi
\expandafter\ifx\csname bibnamefont\endcsname\relax
  \def\bibnamefont#1{#1}\fi
\expandafter\ifx\csname bibfnamefont\endcsname\relax
  \def\bibfnamefont#1{#1}\fi
\expandafter\ifx\csname citenamefont\endcsname\relax
  \def\citenamefont#1{#1}\fi
\expandafter\ifx\csname url\endcsname\relax
  \def\url#1{\texttt{#1}}\fi
\expandafter\ifx\csname urlprefix\endcsname\relax\def\urlprefix{URL }\fi
\providecommand{\bibinfo}[2]{#2}
\providecommand{\eprint}[2][]{\url{#2}}

\bibitem[{\citenamefont{Hohenberg and Kohn}(1964)}]{dft}
\bibinfo{author}{\bibfnamefont{P.}~\bibnamefont{Hohenberg}} \bibnamefont{and}
  \bibinfo{author}{\bibfnamefont{W.}~\bibnamefont{Kohn}},
  \bibinfo{journal}{Phys. Rev.} \textbf{\bibinfo{volume}{136}},
  \bibinfo{pages}{B864} (\bibinfo{year}{1964}).

\bibitem[{\citenamefont{Baettig and Spaldin}(2005)}]{apl_design}
\bibinfo{author}{\bibfnamefont{P.}~\bibnamefont{Baettig}} \bibnamefont{and}
  \bibinfo{author}{\bibfnamefont{N.}~\bibnamefont{Spaldin}},
  \bibinfo{journal}{Appl. Phys. Lett.} \textbf{\bibinfo{volume}{86}},
  \bibinfo{pages}{012505} (\bibinfo{year}{2005}).

\bibitem[{\citenamefont{Tokura}(2000)}]{review_tmo}
\bibinfo{author}{\bibfnamefont{Y.}~\bibnamefont{Tokura}},
  \emph{\bibinfo{title}{Colossal magnetoresistive oxides}}
  (\bibinfo{address}{Australia: Gordon and Breach Science Publishers},
  \bibinfo{year}{2000}).

\bibitem[{\citenamefont{Terasaki et~al.}(1997)\citenamefont{Terasaki, Sasago,
  and Uchinokura}}]{thermo}
\bibinfo{author}{\bibfnamefont{I.}~\bibnamefont{Terasaki}},
  \bibinfo{author}{\bibfnamefont{Y.}~\bibnamefont{Sasago}}, \bibnamefont{and}
  \bibinfo{author}{\bibfnamefont{K.}~\bibnamefont{Uchinokura}},
  \bibinfo{journal}{Phys. Rev. B} \textbf{\bibinfo{volume}{56}},
  \bibinfo{pages}{R12685} (\bibinfo{year}{1997}).

\bibitem[{\citenamefont{Martin et~al.}(1997)\citenamefont{Martin, Maignan,
  Pelloquin, Nguygen, and Raveau}}]{co_cmr}
\bibinfo{author}{\bibfnamefont{C.}~\bibnamefont{Martin}},
  \bibinfo{author}{\bibfnamefont{A.}~\bibnamefont{Maignan}},
  \bibinfo{author}{\bibfnamefont{D.}~\bibnamefont{Pelloquin}},
  \bibinfo{author}{\bibfnamefont{N.}~\bibnamefont{Nguygen}}, \bibnamefont{and}
  \bibinfo{author}{\bibfnamefont{B.}~\bibnamefont{Raveau}},
  \bibinfo{journal}{Appl. Phys. Lett.} \textbf{\bibinfo{volume}{71}},
  \bibinfo{pages}{1421} (\bibinfo{year}{1997}).

\bibitem[{\citenamefont{Takada et~al.}(2003)\citenamefont{Takada, Sakurai,
  Takayama-Muromachi, Izumi, Dilanian, and Sasaki}}]{co_sc}
\bibinfo{author}{\bibfnamefont{K.}~\bibnamefont{Takada}},
  \bibinfo{author}{\bibfnamefont{H.}~\bibnamefont{Sakurai}},
  \bibinfo{author}{\bibfnamefont{E.}~\bibnamefont{Takayama-Muromachi}},
  \bibinfo{author}{\bibfnamefont{F.}~\bibnamefont{Izumi}},
  \bibinfo{author}{\bibfnamefont{R.}{A.}~\bibnamefont{Dilanian}}, \bibnamefont{and}
  \bibinfo{author}{\bibfnamefont{T.}~\bibnamefont{Sasaki}},
  \bibinfo{journal}{Nature (London)} \textbf{\bibinfo{volume}{422}},
  \bibinfo{pages}{53} (\bibinfo{year}{2003}).

\bibitem[{\citenamefont{Anisimov et~al.}(1997)\citenamefont{Anisimov,
  Aryasetiawan, and Lichtenstein}}]{anisimov}
\bibinfo{author}{\bibfnamefont{V.}{I.}~\bibnamefont{Anisimov}},
  \bibinfo{author}{\bibfnamefont{F.}~\bibnamefont{Aryasetiawan}},
  \bibnamefont{and}
  \bibinfo{author}{\bibfnamefont{A.}{I.}~\bibnamefont{Lichtenstein}},
  \bibinfo{journal}{J.\ Phys.: Condens. Matter} \textbf{\bibinfo{volume}{9}},
  \bibinfo{pages}{767} (\bibinfo{year}{1997}).

\bibitem[{\citenamefont{Anisimov et~al.}(1997)\citenamefont{Anisimov, Poteryaev,
  Korotin, Anokhin and Kotliar}}]{dmft}
\bibinfo{author}{\bibfnamefont{V.}{I.}~\bibnamefont{Anisimov}},
  \bibinfo{author}{\bibfnamefont{A.}{I.}~\bibnamefont{Poteryaev}},
  \bibinfo{author}{\bibfnamefont{M.}{A.}~\bibnamefont{Korotin}},
  \bibinfo{author}{\bibfnamefont{A.}{O.}~\bibnamefont{Anokhin}},
  \bibnamefont{and}
  \bibinfo{author}{\bibfnamefont{G.}~\bibnamefont{Kotliar}},
  \bibinfo{journal}{J.\ Phys.: Condens. Matter} \textbf{\bibinfo{volume}{9}},
  \bibinfo{pages}{7359} (\bibinfo{year}{1997}).

\bibitem[{\citenamefont{Wohlfarth}(2000)}]{review_batlle}
\bibinfo{author}{\bibfnamefont{E.}{P.}~\bibnamefont{Wohlfarth}},
  \bibinfo{journal}{J. Phys. F: Metal Phys.} \textbf{\bibinfo{volume}{10}},
  \bibinfo{pages}{L241} (\bibinfo{year}{1980}).

\bibitem[{\citenamefont{Mart\'{\i}n et~al.}(2003)\citenamefont{Mart\'{\i}n, 
  Nogu\'es, Liu, Vicent, and Schuller}}]{applications}
\bibinfo{author}{\bibfnamefont{J.}{I.}~\bibnamefont{Mart\'{\i}n}},
  \bibinfo{author}{\bibfnamefont{J.}~\bibnamefont{Nogu\'es}},
  \bibinfo{author}{\bibfnamefont{K.}~\bibnamefont{Liu}},
  \bibinfo{author}{\bibfnamefont{J.}{L.}~\bibnamefont{Vicent}}, \bibnamefont{and}
  \bibinfo{author}{\bibfnamefont{I.}{K.}~\bibnamefont{Schuller}},
  \bibinfo{journal}{J. Magn. Magn. Mater.} \textbf{\bibinfo{volume}{256}},
  \bibinfo{pages}{449} (\bibinfo{year}{2003}).

\bibitem[{\citenamefont{Gao et~al.}(2004)\citenamefont{Gao, Bao, Beerman,
  Yasuhara, Shindo, and Krishnan}}]{apl_coparticles}
\bibinfo{author}{\bibfnamefont{Y.}~\bibnamefont{Gao}},
  \bibinfo{author}{\bibfnamefont{Y.}~\bibnamefont{Bao}},
  \bibinfo{author}{\bibfnamefont{M.}~\bibnamefont{Beerman}},
  \bibinfo{author}{\bibfnamefont{A.}~\bibnamefont{Yasuhara}},
  \bibinfo{author}{\bibfnamefont{D.}~\bibnamefont{Shindo}}, \bibnamefont{and}
  \bibinfo{author}{\bibfnamefont{K.}{M.}~\bibnamefont{Krishnan}},
  \bibinfo{journal}{Appl. Phys. Lett.} \textbf{\bibinfo{volume}{84}},
  \bibinfo{pages}{3361} (\bibinfo{year}{2004}).

\bibitem[{\citenamefont{Pardo et~al.}(2004{\natexlab{a}})\citenamefont{Pardo,
  Rivas, Baldomir, Iglesias, Blaha, Schwarz, and Arias}}]{vpardo_clusters}
\bibinfo{author}{\bibfnamefont{V.}~\bibnamefont{Pardo}},
  \bibinfo{author}{\bibfnamefont{J.}~\bibnamefont{Rivas}},
  \bibinfo{author}{\bibfnamefont{D.}~\bibnamefont{Baldomir}},
  \bibinfo{author}{\bibfnamefont{M.}~\bibnamefont{Iglesias}},
  \bibinfo{author}{\bibfnamefont{P.}~\bibnamefont{Blaha}},
  \bibinfo{author}{\bibfnamefont{K.}~\bibnamefont{Schwarz}}, \bibnamefont{and}
  \bibinfo{author}{\bibfnamefont{J.}{E.}~\bibnamefont{Arias}},
  \bibinfo{journal}{Phys. Rev. B} \textbf{\bibinfo{volume}{70}},
  \bibinfo{pages}{212404} (\bibinfo{year}{2004}{\natexlab{a}}).

\bibitem[{\citenamefont{Se\~nar\'{\i}s Rodr\'{\i}guez and Goodenough}(1995)}]{tona}
\bibinfo{author}{\bibfnamefont{M.}{A.}~\bibnamefont{Se\~nar\'{\i}s
Rodr\'{\i}guez}},
  \bibnamefont{and}
  \bibinfo{author}{\bibfnamefont{J.}{B.}\bibnamefont{Goodenough}},
  \bibinfo{journal}{J. Solid State Chem.} \textbf{\bibinfo{volume}{118}},
  \bibinfo{pages}{323} (\bibinfo{year}{1995}).

\bibitem[{\citenamefont{Mira et~al.}(1997)\citenamefont{Mira, Rivas, S\'anchez,
  nar\'{\i}s Rodr\'{\i}guez, Fiorani, Rinaldi, and Caciuffo}}]{mira97}
\bibinfo{author}{\bibfnamefont{J.}~\bibnamefont{Mira}},
  \bibinfo{author}{\bibfnamefont{J.}~\bibnamefont{Rivas}},
  \bibinfo{author}{\bibfnamefont{R.}{D.}~\bibnamefont{S\'anchez}},
  \bibinfo{author}{\bibfnamefont{M.}{A.}~ \bibnamefont{Se\~nar\'{\i}s
  Rodr\'{\i}guez}}, \bibinfo{author}{\bibfnamefont{D.}~\bibnamefont{Fiorani}},
  \bibinfo{author}{\bibfnamefont{D.}~\bibnamefont{Rinaldi}}, \bibnamefont{and}
  \bibinfo{author}{\bibfnamefont{R.}~\bibnamefont{Caciuffo}},
  \bibinfo{journal}{J. Appl. Phys.} \textbf{\bibinfo{volume}{81}},
  \bibinfo{pages}{5753} (\bibinfo{year}{1997}).

\bibitem[{\citenamefont{Mira et~al.}(1999)\citenamefont{Mira, Rivas, Jonason,
  Nordblad, Breijo, and Se\~nar\'{\i}s Rodr\'{\i}guez}}]{spinglass}
\bibinfo{author}{\bibfnamefont{J.}~\bibnamefont{Mira}},
  \bibinfo{author}{\bibfnamefont{J.}~\bibnamefont{Rivas}},
  \bibinfo{author}{\bibfnamefont{K.}~\bibnamefont{Jonason}},
  \bibinfo{author}{\bibfnamefont{P.}~\bibnamefont{Nordblad}},
  \bibinfo{author}{\bibfnamefont{M.}{P.}~\bibnamefont{Breijo}}, \bibnamefont{and}
  \bibinfo{author}{\bibfnamefont{M.}{A.}~\bibnamefont{Se\~nar\'{\i}s
  Rodr\'{\i}guez}}, \bibinfo{journal}{J. Magn. Magn. Mater.}
  \textbf{\bibinfo{volume}{196-197}}, \bibinfo{pages}{487}
  (\bibinfo{year}{1999}).

\bibitem[{\citenamefont{Mira et~al.}(2001)\citenamefont{Mira, Rivas, Baio,
Barucca,
  Caciuffo, Rinaldi, Fiorani, and Se\~nar\'{\i}s Rodr\'{\i}guez}}]{mira01}
\bibinfo{author}{\bibfnamefont{J.}~\bibnamefont{Mira}},
  \bibinfo{author}{\bibfnamefont{J.}~\bibnamefont{Rivas}},
  \bibinfo{author}{\bibfnamefont{G.}~\bibnamefont{Baio}},
  \bibinfo{author}{\bibfnamefont{G.}~\bibnamefont{Barucca}},
  \bibinfo{author}{\bibfnamefont{R.}~\bibnamefont{Caciuffo}},
  \bibinfo{author}{\bibfnamefont{D.}~\bibnamefont{Rinaldi}},
  \bibinfo{author}{\bibfnamefont{D.}~\bibnamefont{Fiorani}}, \bibnamefont{and}
  \bibinfo{author}{\bibfnamefont{M.}{A.}~\bibnamefont{Se\~nar\'{\i}s
  Rodr\'{\i}guez}}, \bibinfo{journal}{J. Appl. Phys.}
  \textbf{\bibinfo{volume}{89}}, \bibinfo{pages}{5606} (\bibinfo{year}{2001}).

\bibitem[{\citenamefont{Szymczak et~al.}(2005)\citenamefont{Szymczak, Baran,
  Babonas, Diduszko, Fink-Finowicki, and Szymczak}}]{jmmm_glass}
\bibinfo{author}{\bibfnamefont{H.}~\bibnamefont{Szymczak}},
  \bibinfo{author}{\bibfnamefont{M.}~\bibnamefont{Baran}},
  \bibinfo{author}{\bibfnamefont{G.}{J.}~\bibnamefont{Babonas}},
  \bibinfo{author}{\bibfnamefont{R.}~\bibnamefont{Diduszko}},
  \bibinfo{author}{\bibfnamefont{J.}~\bibnamefont{Fink-Finowicki}},
  \bibnamefont{and} \bibinfo{author}{\bibfnamefont{R.}~\bibnamefont{Szymczak}},
  \bibinfo{journal}{J. Magn. Magn. Mater.} \textbf{\bibinfo{volume}{285}},
  \bibinfo{pages}{386} (\bibinfo{year}{2005}).

\bibitem[{\citenamefont{Rivas et~al.}(2005)\citenamefont{Rivas, Mira, Rinaldi,
  Caciuffo, and Se\~nar\'{\i} Rodr\'{\i}uez}}]{rivas}
\bibinfo{author}{\bibfnamefont{J.}~\bibnamefont{Rivas}},
  \bibinfo{author}{\bibfnamefont{J.}~\bibnamefont{Mira}},
  \bibinfo{author}{\bibfnamefont{D.}~\bibnamefont{Rinaldi}},
  \bibinfo{author}{\bibfnamefont{R.}~\bibnamefont{Caciuffo}},
  \bibnamefont{and} 
  \bibinfo{author}{\bibfnamefont{M.}{A.}~\bibnamefont{Se\~nar\'{\i} 
  Rodr\'{\i}uez}}, \bibinfo{journal}{J. Magn. Magn.
  Mater.} \textbf{\bibinfo{volume}{in press}} (\bibinfo{year}{2005}).

\bibitem[{\citenamefont{Rivadulla et~al.}(2004)\citenamefont{Rivadulla,
  L\'opez-Quintela, and Rivas}}]{rivadulla_prl}
\bibinfo{author}{\bibfnamefont{F.}~\bibnamefont{Rivadulla}},
  \bibinfo{author}{\bibfnamefont{M.}{A.}~\bibnamefont{L\'opez-Quintela}},
  \bibnamefont{and} \bibinfo{author}{\bibfnamefont{J.}~\bibnamefont{Rivas}},
  \bibinfo{journal}{Phys. Rev. Lett.} \textbf{\bibinfo{volume}{93}},
  \bibinfo{pages}{167206} (\bibinfo{year}{2004}).

\bibitem[{\citenamefont{Rivas et~al.}(2004)\citenamefont{Rivas, Rivadulla, and
  L\'opez-Quintela}}]{rivas_physb}
\bibinfo{author}{\bibfnamefont{J.}~\bibnamefont{Rivas}},
  \bibinfo{author}{\bibfnamefont{F.}~\bibnamefont{Rivadulla}},
  \bibnamefont{and}
  \bibinfo{author}{\bibfnamefont{M.}{A.}~\bibnamefont{L\'opez-Quintela}},
  \bibinfo{journal}{Phys. B.} \textbf{\bibinfo{volume}{354}},
  \bibinfo{pages}{1} (\bibinfo{year}{2004}).

\bibitem[{\citenamefont{Pardo et~al.}(2004{\natexlab{b}})\citenamefont{Pardo,
  Blaha, Iglesias, Schwarz, Baldomir, and Arias}}]{vpardo_prb}
\bibinfo{author}{\bibfnamefont{V.}~\bibnamefont{Pardo}},
  \bibinfo{author}{\bibfnamefont{P.}~\bibnamefont{Blaha}},
  \bibinfo{author}{\bibfnamefont{M.}~\bibnamefont{Iglesias}},
  \bibinfo{author}{\bibfnamefont{K.}~\bibnamefont{Schwarz}},
  \bibinfo{author}{\bibfnamefont{D.}~\bibnamefont{Baldomir}}, \bibnamefont{and}
  \bibinfo{author}{\bibfnamefont{J.}{E.}~\bibnamefont{Arias}},
  \bibinfo{journal}{Phys. Rev. B} \textbf{\bibinfo{volume}{70}},
  \bibinfo{pages}{144422} (\bibinfo{year}{2004}{\natexlab{b}}).

\bibitem[{\citenamefont{Yamaura et~al.}(1999)\citenamefont{Yamaura, Zandbergen,
  Abe, and Cava}}]{yamaura}
\bibinfo{author}{\bibfnamefont{K.}~\bibnamefont{Yamaura}},
  \bibinfo{author}{\bibfnamefont{H.}{W.}~\bibnamefont{Zandbergen}},
  \bibinfo{author}{\bibfnamefont{K.}~\bibnamefont{Abe}}, \bibnamefont{and}
  \bibinfo{author}{\bibfnamefont{R.}{J.}~\bibnamefont{Cava}}, \bibinfo{journal}{J.
  Solid State Chem.} \textbf{\bibinfo{volume}{146}}, \bibinfo{pages}{96}
  (\bibinfo{year}{1999}).

\bibitem[{\citenamefont{Blanco-Mantec\'on and O'Grady}(1999)}]{bimodality}
\bibinfo{author}{\bibfnamefont{M.}~\bibnamefont{Blanco-Mantec\'on}}
  \bibnamefont{and} \bibinfo{author}{\bibfnamefont{K.}~\bibnamefont{O'Grady}},
  \bibinfo{journal}{J. Magn. Magn. Mater.} \textbf{\bibinfo{volume}{203}},
  \bibinfo{pages}{50} (\bibinfo{year}{1999}).

\bibitem[{\citenamefont{De Biasi et~al.}(2002)\citenamefont{De Biasi, Ramos, Zysler,
  and Romero}}]{prb_surface}
\bibinfo{author}{\bibfnamefont{E.}~\bibnamefont{De Biasi}},
  \bibinfo{author}{\bibfnamefont{C.}{A.}~\bibnamefont{Ramos}},
  \bibinfo{author}{\bibfnamefont{R.}{D.}~\bibnamefont{Zysler}}, \bibnamefont{and}
  \bibinfo{author}{\bibfnamefont{H.}~\bibnamefont{Romero}},
  \bibinfo{journal}{Phys. Rev. B} \textbf{\bibinfo{volume}{65}},
  \bibinfo{pages}{144416} (\bibinfo{year}{2002}).

\bibitem[{\citenamefont{Tronc et~al.}(2003)\citenamefont{Tronc, Fiorani,
  Nogu\`{e}s, Testa, Lucari \textsl{et al.}}}]{fiorani_surf}
\bibinfo{author}{\bibfnamefont{E.}~\bibnamefont{Tronc}},
  \bibinfo{author}{\bibfnamefont{D.}~\bibnamefont{Fiorani}},
  \bibinfo{author}{\bibfnamefont{M.}~\bibnamefont{Nogu\`{e}s}},
  \bibinfo{author}{\bibfnamefont{A.}{M.}~\bibnamefont{Testa}},
  \bibinfo{author}{\bibfnamefont{F.}~\bibnamefont{Lucari},
  \bibnamefont{\textsl{et al.}}}, \bibinfo{journal}{J. Magn. Magn. Mater.}
  \textbf{\bibinfo{volume}{262}}, \bibinfo{pages}{6} (\bibinfo{year}{2003}).

\bibitem[{\citenamefont{Schwarz and Blaha}(2003)}]{wien}
\bibinfo{author}{\bibfnamefont{K.}~\bibnamefont{Schwarz}} \bibnamefont{and}
  \bibinfo{author}{\bibfnamefont{P.}~\bibnamefont{Blaha}},
  \bibinfo{journal}{Comp. Mat. Sci.} \textbf{\bibinfo{volume}{28}},
  \bibinfo{pages}{259} (\bibinfo{year}{2003}).

\bibitem[{\citenamefont{Sj{\"o}stedt et~al.}(2000)\citenamefont{Sj{\"o}stedt,
  N{\"o}rdstrom, and Singh}}]{sjo}
\bibinfo{author}{\bibfnamefont{E.}~\bibnamefont{Sj{\"o}stedt}},
  \bibinfo{author}{\bibfnamefont{L.}~\bibnamefont{N{\"o}rdstrom}},
  \bibnamefont{and} \bibinfo{author}{\bibfnamefont{D.}{J.}~\bibnamefont{Singh}},
  \bibinfo{journal}{Solid State Commun.} \textbf{\bibinfo{volume}{114}},
  \bibinfo{pages}{15} (\bibinfo{year}{2000}).

\bibitem[{\citenamefont{Lichtenstein et~al.}(1995)\citenamefont{Lichtenstein,
  Anisimov, and Zaanen}}]{sic}
\bibinfo{author}{\bibfnamefont{A.}{I.}~\bibnamefont{Lichtenstein}},
  \bibinfo{author}{\bibfnamefont{V.}{I.}~\bibnamefont{Anisimov}}, \bibnamefont{and}
  \bibinfo{author}{\bibfnamefont{J.}~\bibnamefont{Zaanen}},
  \bibinfo{journal}{Phys.\ Rev. B} \textbf{\bibinfo{volume}{52}},
  \bibinfo{pages}{R5467} (\bibinfo{year}{1995}).

\bibitem[{\citenamefont{Petukhov et~al.}(2003)\citenamefont{Petukhov, Mazin,
  Chioncel, and Lichtenstein}}]{mazin}
\bibinfo{author}{\bibfnamefont{A.}{G.}~\bibnamefont{Petukhov}},
  \bibinfo{author}{\bibfnamefont{I.}{I.}~\bibnamefont{Mazin}},
  \bibinfo{author}{\bibfnamefont{L.}~\bibnamefont{Chioncel}}, \bibnamefont{and}
  \bibinfo{author}{\bibfnamefont{A.}{I.}~\bibnamefont{Lichtenstein}},
  \bibinfo{journal}{Phys.\ Rev. B} \textbf{\bibinfo{volume}{67}},
  \bibinfo{pages}{153106} (\bibinfo{year}{2003}).

\bibitem[{\citenamefont{Singh}(1994)}]{singh}
\bibinfo{author}{\bibfnamefont{D.}{J.}~\bibnamefont{Singh}},
  \emph{\bibinfo{title}{Planewaves, pseudopotentials and LAPW method}}
  (\bibinfo{address}{Kluwer Academic Publishers}, \bibinfo{year}{1994}).

\bibitem[{\citenamefont{Morrish}(2001)}]{morrish}
\bibinfo{author}{\bibfnamefont{A.}{H.}~\bibnamefont{Morrish}},
  \emph{\bibinfo{title}{The Physical Principles of Magnetism}}
  (\bibinfo{address}{IEEE Press, New York}, \bibinfo{year}{2001}).

\bibitem[{\citenamefont{Binder and Young}(1986)}]{review_spinglass}
\bibinfo{author}{\bibfnamefont{K.}~\bibnamefont{Binder}} \bibnamefont{and}
  \bibinfo{author}{\bibfnamefont{A.}{P.}~\bibnamefont{Young}},
  \bibinfo{journal}{Rev. Mod. Phys.} \textbf{\bibinfo{volume}{58}},
  \bibinfo{pages}{801} (\bibinfo{year}{1986}).

\bibitem[{\citenamefont{Dormann et~al.}(1997)\citenamefont{Dormann, Fiorani,
  and Tronc}}]{fiorani}
\bibinfo{author}{\bibfnamefont{J.}{L.}~\bibnamefont{Dormann}},
  \bibinfo{author}{\bibfnamefont{D.}~\bibnamefont{Fiorani}}, \bibnamefont{and}
  \bibinfo{author}{\bibfnamefont{E.}~\bibnamefont{Tronc}},
  \bibinfo{journal}{Adv. Chem. Phys.} \textbf{\bibinfo{volume}{98}},
  \bibinfo{pages}{283} (\bibinfo{year}{1997}).

\bibitem[{\citenamefont{Garc\'{\i}a-Otero
  et~al.}(2000)\citenamefont{Garc\'{\i}a-Otero, Porto, Rivas, and
  Bunde}}]{cooling_speed}
\bibinfo{author}{\bibfnamefont{J.}~\bibnamefont{Garc\'{\i}a-Otero}},
  \bibinfo{author}{\bibfnamefont{M.}~\bibnamefont{Porto}},
  \bibinfo{author}{\bibfnamefont{J.}~\bibnamefont{Rivas}}, \bibnamefont{and}
  \bibinfo{author}{\bibfnamefont{A.}~\bibnamefont{Bunde}},
  \bibinfo{journal}{Phys. Rev. Lett.} \textbf{\bibinfo{volume}{84}},
  \bibinfo{pages}{167} (\bibinfo{year}{2000}).

\end{thebibliography}

\end{document}